\documentclass[aps,prl,twocolumn,groupedaddress]{revtex4-1}
\usepackage{amsmath}
\usepackage{parskip}
\usepackage{enumitem}
\usepackage{amssymb}
\usepackage{overpic}
\usepackage{hyperref}
\usepackage{soul}
\usepackage[makeroom]{cancel}

\usepackage{float,graphicx}
\usepackage[usenames, dvipsnames]{color}

\hypersetup{
	colorlinks,
	citecolor=blue,
	filecolor=red,
	linkcolor=blue,
	urlcolor=blue,
	hyperfigures
}

\def\Xint#1{\mathchoice
   {\XXint\displaystyle\textstyle{#1}}%
   {\XXint\textstyle\scriptstyle{#1}}%
   {\XXint\scriptstyle\scriptscriptstyle{#1}}%
   {\XXint\scriptscriptstyle\scriptscriptstyle{#1}}%
   \!\int}
\def\XXint#1#2#3{{\setbox0=\hbox{$#1{#2#3}{\int}$}
     \vcenter{\hbox{$#2#3$}}\kern-.5\wd0}}

\def\dashint{\Xint-}

\usepackage[page]{appendix}

\graphicspath{{./Figures/}}

\newcommand{\bff}{{\bf f}}
\newcommand{\bu}{{\bf u}}

\newcommand{\bx}{{\bf x}}

\newcommand{\pll}{\parallel}
\newcommand{\uvc}[1]{\boldsymbol{\mathrm{\hat #1}}}

\begin{document}

\title{Hydrodynamic synchronization of spontaneously beating filaments}
\author{Brato Chakrabarti}
\author{David Saintillan}
\email[Email address: ]{dstn@ucsd.edu}
\affiliation{Department of Mechanical and Aerospace Engineering, University of California San Diego, 9500 Gilman Drive, La Jolla, CA 92093, USA}

\date{\today}

\begin{abstract}
{Using a  {geometric feedback} model of the flagellar axoneme accounting for dynein motor kinetics, we study elastohydrodynamic phase synchronization in a pair of spontaneously beating filaments with waveforms ranging from sperm to cilia and \textit{Chlamydomonas}. Our computations reveal that both in-phase and anti-phase synchrony can emerge for asymmetric beats while   {symmetric waveforms} go in-phase, and elucidate the mechanism for phase slips due to biochemical noise.\ Model predictions agree with recent experiments and illuminate the crucial roles of hydrodynamics and mechanochemical feedback in synchronization.}	
\end{abstract}
\pacs{...}
\maketitle

Studies on  {flagellar} synchronization date back to observations  by Rothschild \cite{rothschild1949measurement} on nearby swimming sperms and subsequent theoretical work by Taylor \cite{taylor1951analysis}, who proved that dissipation for two swimming sheets is minimized for an in-phase configuration.\ While biology is often not driven by dissipation principles, it has long been hypothesized that hydrodynamic interactions play a central role in synchronization \cite{elfring2009hydrodynamic} and in collective behaviors such as metachronal waves in ciliary arrays \cite{golestanian2011hydrodynamic}.\ Over the last two decades, experiments \cite{ruffer1985high,polin2009chlamydomonas,goldstein2009noise,leptos2013antiphase,wan2014lag,brumley2014flagellar} using micropipette-held \textit{Chlamydomonas} have revealed that elastohydrodynamic interactions may indeed be at play in causing its two flagella to synchronize their breaststrokes, with periods of asynchrony thought to arise due to biochemical noise  \cite{brumley2014flagellar,pikovsky2003synchronization}.\ 

Theoretical progress in understanding synchronization is complicated by the intricate internal structure and actuation of the flagellum core, or axoneme. In presence of ATP, thousands of dynein molecular motors act in concert to bend the structure and drive spontaneous beats.\ Much work has gone into developing minimal models that neglect this biological complexity and coarse-grain flagella as microspheres driven on compliant or  tilted orbits \cite{vilfan2006hydrodynamic,niedermayer2008synchronization,uchida2011generic,maestro2018control}.\ More detailed numerical models have relied on pre-imposed internal or external actuations to analyze metachronal waves \cite{gueron1997cilia,elgeti2013emergence} or the bistability of elastic filaments \cite{guo2018bistability}, yet these descriptions poorly capture experimental waveforms \cite{Note1}.\ Only recently have there been attempts to study the role of hydrodynamics in simplified models of active  microfilaments~\cite{goldstein2016elastohydrodynamic}. 

While the detailed process leading to spontaneous flagellar oscillations remains controversial, several mechanisms have been proposed ranging from flutter-like instabilities \cite{bayly2016steady,de2017spontaneous,ling2018instability} to dynamic internal tension \cite{han2018spontaneous} and geometric control of dynein kinetics \cite{lindemann1994geometric,brokaw1971bend,bayly2014equations,bayly2015analysis,sartori2016dynamic,oriola2017nonlinear,sartori2019effect}.\  {The role of dynein motors in driving oscillations of microtubules has also been established in purified \textit{in vitro} systems \cite{vilfan2019}.}
Building on previous sliding  {and curvature} control models \cite{riedel2007molecular,oriola2017nonlinear,sartori2016dynamic,sartori2019effect}, we recently developed \cite{chakrabartispontaneous} a microscopic description for an active elastic flagellum accounting for internal dynein motor kinetics, which produces spontaneous oscillations following saturation of a Hopf bifurcation and can generate a variety of beating patterns observed in nature.\ In this Letter, this microscopic model is employed to analyze the temporal dynamics and synchronization of a pair of spontaneously beating filaments.\ Our results  {explain synchronization in various situations with trends consistent with experiments \cite{brumley2014flagellar}, and underscore} the crucial roles of hydrodynamic interactions, mechanochemical feedback, and biochemical noise.

\begin{figure*}[t]
	\includegraphics[width=\textwidth]{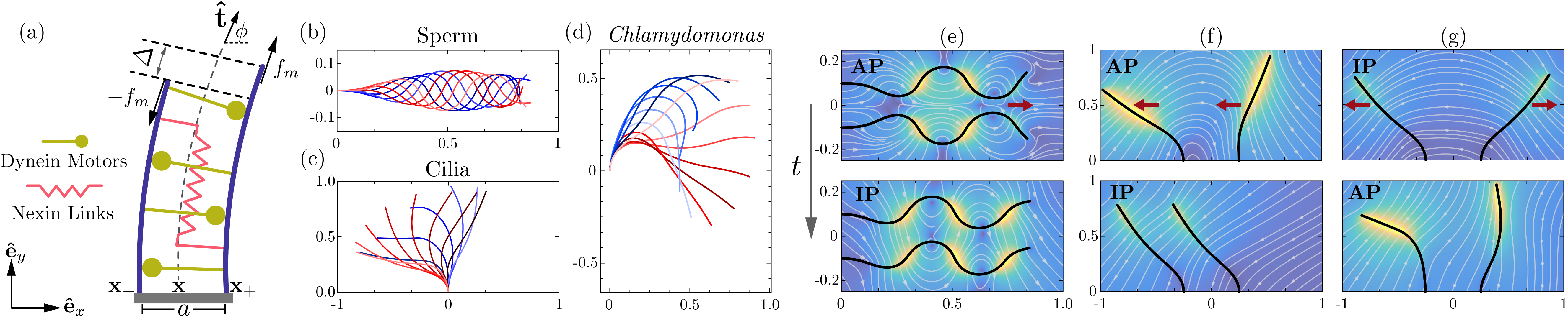}
	\caption{(a) Schematic representation of the planar model for the flagellar axoneme.\ (b)--(d) Spontaneous beating patterns emerging from the nonlinear model that approximate the waveforms of sperm, cilia and \textit{Chlamydomonas}. (e)--(g) Synchronization of different beating patterns.\ The top panel shows snapshots at $t=0$ and the bottom panel illustrates the final configurations.\ Sperms (e) beat in phase (IP), while cilia (f)--(g) can achieve both IP or AP synchronization depending on the orientation of the power stroke indicated by red arrows. Synchronization for \emph{Chlamydomonas} (not shown) is identical to that of cilia.\ See movies showing the temporal dynamics in \cite{Note1}.}
	\label{fig:Fig1}
	\vspace*{-0.em}
\end{figure*}

\textit{Model formulation.---}The flagellar axoneme is a 3D structure with circular cross-section composed of 9+2 pairs of microtubules arranged in a cyclic fashion \cite{alberts2013essential}.\ Following past models \cite{machin1958wave,brokaw1965non,brokaw1971bend,brokaw1979calcium,oriola2017nonlinear,chakrabartispontaneous,camalet2000generic}, we idealize this structure as a planar projection with diameter $a$ and length $L$, where microtubules are represented by two polar filaments $\bx_{\pm}$ clamped at the base [Fig.\ \ref{fig:Fig1}(a)].\ We seek an evolution equation for the centerline $\bx(s,t)$ parametrized by arclength $s\in[0,L]$.\ Internal actuation arises from dynein motors that extend from each filament and stochastically bind with the opposite one.\ In presence of ATP, these motors move along the filaments and generate internal shear forces resulting  in an arclength mismatch known as the sliding displacement \cite{camalet2000generic}: $\smash{\Delta(s,t) = \int_0^s \left(|\partial_{s'} \bx_-| - |\partial_{s'} \bx_+|\right) ds' = a \phi(s,t)}$, where ${\phi}(s,t)$ is the tangent angle and  {$(\cdot)_s$ denotes arclength derivative}.\ This sliding is resisted by internal protein linkers, or nexin links, modeled as linear springs of stiffness $K$.\ Both dynein motors and nexin links result in equal and opposite force densities along the filaments:
\begin{equation}\label{eq:slideforce}
f_m(s,t) = \rho\, (n_+ F_+  + n_-F_-) - K \Delta,
\end{equation}
where $\rho$ is the mean motor density, $n_{\pm}$ are the fractions of motors in the bound state, and $F_{\pm}$ are the associated loads.\ These sliding forces generate internal moments $\smash{M(s,t) = B \phi_s - a \int_s^L f_m(s',t) ds'}$, where the first term captures the passive elastic response of the structure modeled as an inextensible Euler-Bernoulli elastica with bending rigidity $B$ \cite{duroure2019}. Dynein motor populations evolve as $\partial_t n_\pm = \pi_\pm - \epsilon_\pm$ where $\pi_{\pm}$ and $\epsilon_{\pm}$ are the attachment and detachment rates, respectively.\ The attachment rate is proportional to the fraction of unbound motors: $\pi_{\pm} = \pi_0 \left(1-n_{\pm}\right)$.\ The detachment rate depends linearly on the fraction of bound motors and exponentially on the carried load \cite{svoboda94,muller2008tug}:~$\epsilon_{\pm} = \epsilon_0 n_\pm \exp(\pm F_\pm/f_c)$, where $f_c$ is a critical load above which rapid unbinding occurs. To complete the model with the appropriate geometric feedback, we specify a force--velocity relation for the dyneins \cite{oriola2017nonlinear,shelley2016dynamics}: we assume that the motors have a velocity $v_0$ at zero load that decreases linearly with sliding velocity $\Delta_t \equiv a \phi_t$ and are able to carry a load $f_0$ when stalled, yielding the expression: $F_\pm = \pm f_0 (1\mp\Delta_t/v_0)$.

Motion of the centerline is governed by the force balance for an elastic rod in viscous flow \cite{antmannonlinear}: $\partial_s \mathbf{F}_{e} + \bff_{v} = 0$, where $\mathbf{F}_{e}(s,t) = \sigma \uvc{t} + N \uvc{n}$ is the elastic force with tension $\sigma$ and normal force $N$, and $\bff_{v}(s,t)$ is the viscous force density captured by nonlocal slender body theory \cite{keller1976,tornberg2004simulating}.\ This is accompanied by a moment balance in the $z$ direction: $M_s  + N = 0$. Scaling lengths by $L$, sliding displacement by $a$, time by the correlation timescale $\tau_0 =1/(\pi_0 + \epsilon_0)$, elastic forces by $B/L^2$, and motor loads by $\rho f_0$ produces four dimensionless groups, of which two are of primary interest:~(i) the sperm number $\smash{\mathrm{Sp} = L\left(8 \pi \nu/B \tau_0\right)^{1/4}}$, where $\nu$ is the  viscosity, compares the relaxation time of a bending mode to the motor correlation time; (ii) the activity number $\smash{\mu_a = a \rho f_0  L^2/B}$ compares motor-induced sliding forces to characteristic elastic forces.\ The two other dimensionless groups are: $\smash{\mu = K a^2 L^2/B}$ and $\zeta = a/(v_0 \tau_0)$ \cite{chakrabartispontaneous}.\ With these scalings, the dimensionless equations for $\phi(s,t)$ read
\begin{align}
\begin{split}
\sigma_{ss} - \Big(1 + \frac{c_\pll}{c_\perp}\Big) N_s \phi_{s} - N \phi_{ss} -\frac{c_\pll}{c_\perp}\sigma \phi_s^2  & \\
=c_\pll\big(\phi_s u^d_n - &\,\partial_s u^d_t\big), \label{nd1} 
\end{split} \\[6pt]
\begin{split}
N_{ss} - \frac{c_\perp}{c_\pll}  N \phi_s^2 + \sigma \phi_{ss} +  \Big(1 + \frac{c_\perp}{c_\pll}\Big)\sigma_s \phi_s  &\\
=c_\perp\big(\mathrm{Sp}^4 \phi_t - u^d_t \phi_s - &\,\partial_s  u^d_n\big), \label{nd2}
\end{split} 
\end{align}\vspace{-0.4cm}
\begin{align}
\phi_{ss} + \mu_a f_m + N = 0,  \label{nd3}
\end{align}
where the first two equations are force balances in the tangential and normal directions while the third is the moment balance. The dimensionless tangential and normal drag coefficients $c_{\pll,\perp}$ derive from local terms in slender body theory \cite{chakrabartispontaneous} and satisfy $c_\perp/c_\pll \to 2$ for infinitely slender filaments.\ Hydrodynamic interactions are captured by the disturbance velocity $\mathbf{u}^d$, with projections $u_t^d$ and $u_n^d$ in the tangential and normal directions, respectively.\ Given two filaments indexed by $\{\alpha,\beta\}$, the flow is obtained as \vspace{-0.cm}
\begin{equation}
\bu^d(s_\alpha) =  \mathbf{K}[\bff^\alpha_{e}](s_\alpha) + \int_{0}^1 \mathbf{G}(s_\alpha;s_\beta) \cdot \bff^{\beta}_e(s_\beta)\, ds_\beta, \label{eq:flow} \vspace{-0.cm}
\end{equation}
where $\bff_e\equiv\partial_s\mathbf{F}_e$ is the elastic force density.\ The first term in Eq.~\eqref{eq:flow} is the finite-part integral of slender body theory \cite{keller1976,tornberg2004simulating,chakrabartispontaneous} and captures hydrodynamic interactions within a filament.\ The second term accounts for the flow induced by the other filament, with the Green's function $\mathbf{G}(s_\alpha;s_\beta)$ given by the Oseen tensor.\ These equations are supplemented by clamped boundary conditions at $s=0$, and moment- and force-free conditions at $s=1$  \cite{chakrabartispontaneous}. In dimensionless form, the evolution equation for the bound motor populations reads
\begin{align}\label{eq:kinetics}
\begin{split}
\partial_t n_\pm = &\,\eta (1 - n_\pm)  \\&-(1-\eta) n_\pm \exp\left[{f}^* (1 \mp \zeta \Delta_t) \right]+ \xi(s,t),
\end{split}
\end{align}
where $\eta = \pi_0/(\epsilon_0 + \pi_0)$ is the fraction of time spent by motors in the bound state and ${f}^* = f_0/f_c$ is the ratio of the stall load to the characteristic unbinding force.\ The last term accounts for biochemical noise with $\langle\xi(s,t)\rangle=0$ and $\langle\xi(s,t)\xi(s',t')\rangle=2\Lambda \delta(s-s')\delta(t-t')$, where $\Lambda$ is an effective temperature.\ These governing equations are solved numerically as outlined in \cite{chakrabartispontaneous}.

\textit{Spontaneous oscillations.---}We first describe the dynamics of isolated filaments, with model parameters estimated from experiments \cite{riedel2007molecular,oriola2017nonlinear,chakrabartispontaneous};  see Supplemental Information \cite{Note1} for details.\ With a choice of $L \sim 50\, \mu$m for human sperm, $B \sim 0.9 -1.7 \times 10^{-21}\,$N\,m$^2$, $f_0\sim1-5\,$pN, $\tau_0 \sim 50\,$ms and $\rho \sim 10^3\,\mu$m$^{-1}$, we estimate $\mathrm{Sp} \sim 8-20$ and $\mu_a \sim 2-10 \times 10^3$ and explore beating patterns in this range. For a given sperm number, a Hopf bifurcation occurs beyond a critical activity level $\mu_a^c$ and gives rise to spontaneous traveling waves \cite{Note1,chakrabartispontaneous}.\ Close to the bifurcation, the waves propagate from the free end towards the base as previously seen in other simulations of sliding control models  \cite{oriola2017nonlinear,riedel2007molecular}  {but in disagreement with typical sperm beating patterns.}\ However, far from the bifurcation, nonlinearities give rise to a reversal in the direction of propagation \cite{chakrabartispontaneous}, with sperm-like waveforms shown in  Fig.~\ref{fig:Fig1}(b) that resemble experiments \cite{brokaw1965non} and have beating frequencies $f \sim 10-15\,$Hz.\ In the following discussion, we focus on this anterograde propagation regime as it is biologically most relevant.

Asymmetric beating patterns more typical of cilia can be captured by setting different attachment and detachment rates for the motor populations on $\mathbf{x}_\pm$ \cite{chakrabartispontaneous}.\ This bias in the kinetics allows the flagellum to bend in one direction preferentially, resulting in asymmetric power and recovery strokes as shown in Fig.~\ref{fig:Fig1}(c).\ The flagella of wildtype \textit{Chlamydomonas} also have a static mode of deformation \cite{sartori2016dynamic} that we account for using a spontaneous shape $\phi^0(s)$.\ To better approximate their asymmetric breaststrokes [Fig.~\ref{fig:Fig1}(d)], a curvature control mechanism is introduced along with the biased kinetics \cite{chakrabartispontaneous} that uses a generalized Bell's law for the dynein detachment rate: {$\epsilon_\pm = \epsilon_0 n_\pm \exp\left(F_\pm/f_c \pm (\kappa(s)-\phi^0_s(s))/\kappa_c\right)$}, where $\kappa(s)$ is the curvature and $\kappa_c$ is the threshold value for rapid dissociation. {The subtraction of the zero mode in the curvature control follows Sartori et al.~\cite{sartori2016dynamic}, who suggested that motor forces respond to derivatives of curvature rather than curvature itself.}\ Accounting for the short length $L \sim 6-15 \mu$m \cite{sartori2016dynamic} of cilia and \emph{Chlamydomonas} flagella with $B \sim 0.5-5 \times 10^{-22}\,$N\,m$^2$, we estimate $\mathrm{Sp} \sim 2-3$ and measure spontaneous frequencies of $f \sim 10-20$\,Hz.

\begin{figure}[b]
	\centering\vspace{-0.2cm}
		\includegraphics[width=\linewidth]{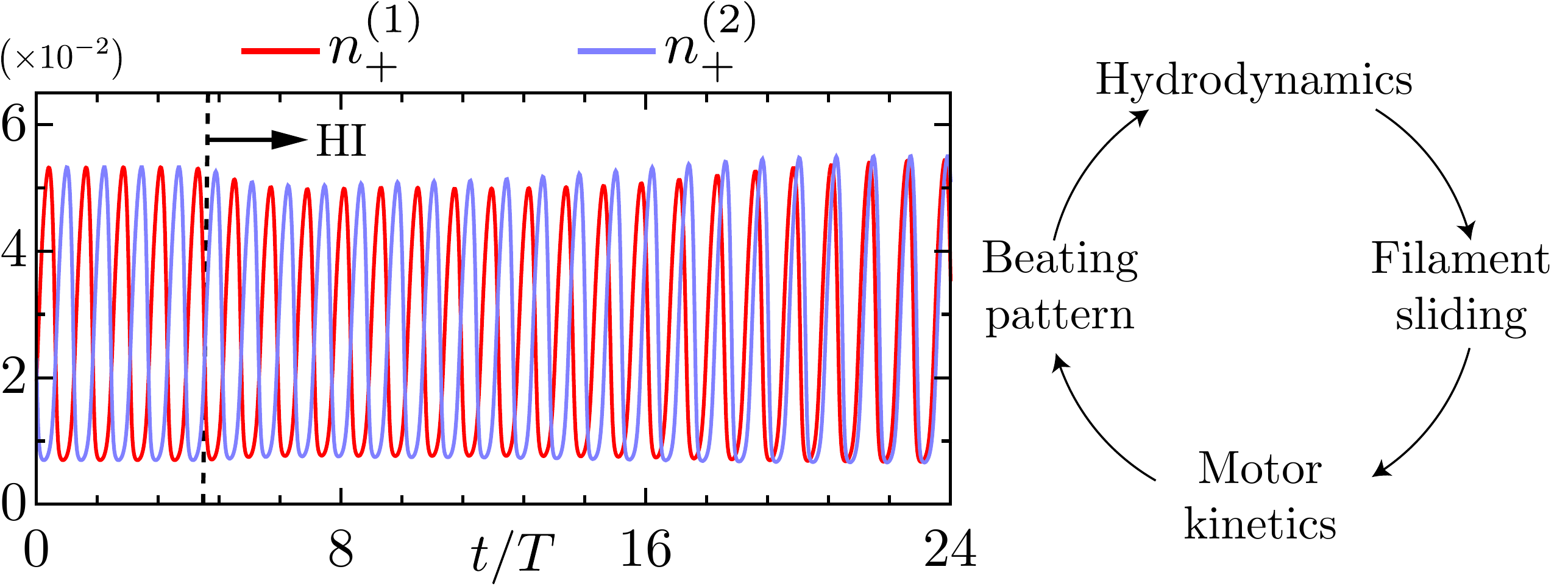}
	\caption{Evolution of dynein motor populations at $s=1/4$ on two nearby sperms as a result of HI. The dashed line indicates the instant when interactions are turned on.\ Time is scaled by the oscillation period $T$ of an isolated filament.\ Cartoon on the right illustrates the feedback loop leading to synchronization.  }
	\label{fig:kinetics}
\end{figure}

\textit{Pair synchronization.---}We first focus on the synchronization of pairs of sperms placed side by side as shown in Fig.~\ref{fig:Fig1}(e).\ We initialize the simulation in absence of inter-filament hydrodynamic interactions (HI) by letting spontaneous oscillations reach steady state after saturation of dynein kinetics.\ The initial configuration is chosen such that the filaments are almost in antiphase (AP) [top panel of Fig.~\ref{fig:Fig1}(e)].\ We then switch on HI and, after several periods, the sperms go in-phase (IP) and remain phase-locked thereafter [bottom panel of Fig.~\ref{fig:Fig1}(e); see movies in \cite{Note1}].\ The key role of hydrodynamics in this process is best illustrated by Fig.~\ref{fig:kinetics}, showing the evolution of the bound motor populations $\smash{n_{+}}$ at $s=1/4$ on both filaments (the behavior is identical for $n_-$ and at other locations).\ Before HI are switched on, motor populations are uncoupled and undergo periodic oscillations in antiphase with cusp-shaped waveforms typical of motors far from equilibrium \cite{julicher1995cooperative} and only a small fraction of bound motors at any given time. Once HI start acting, both the phase and amplitude of the motor populations change. This is attributed to elastic deformations of the filaments in their induced flow fields, which feed back to the kinetics through the change in sliding displacement and velocity.\ As seen in Fig.~\ref{fig:kinetics}, the two motor populations rapidly go in phase with a marginally increased amplitude, resulting in spontaneous IP synchronization of the beating patterns. The cartoon in Fig.~\ref{fig:kinetics} highlights this cyclic process fundamental to elastohydrodynamic synchronization, by which HI affect beating patterns via geometry-dependent motor kinetics.\ This feedback is most dramatic when the filaments are closeby and sufficiently flexible.  

A similar mechanism is at play for asymmetric ciliary beats in Fig.~\ref{fig:Fig1}(f,\,g).\ When the power strokes of the two cilia indicated by red arrows point in the same direction, an IP beat emerges with net unidirectional pumping of the fluid [Fig.~\ref{fig:Fig1}(f)].\ When the power strokes are in opposite directions, our model leads to AP synchronization with beating patterns resembling a `freestyle' swimming gait as shown in Fig.~\ref{fig:Fig1}(g).\ Similar AP patterns are obtained for \textit{Chlamydomonas} beats.  {These observations hint} at the hypothesis \cite{wan2016coordinated} that the IP breaststrokes seen in wildtype cells result from elastic basal couplings between the two flagellar axonemes rather than from HI alone.\ Indeed, experiments with \emph{vfl} mutants that are deficient in these filamentary connections  \cite{wan2016coordinated} or with \emph{Volvox} cells held in separate micropipettes \cite{brumley2014flagellar} have shown AP synchronization for power strokes with opposite orientations, consistent with our model findings.  {Note that in the case of swimming or even weakly clamped cells flagellar synchronization can also happen through a rocking motion of the cell body independent of HI or in absence of basal coupling \cite{friedrich2012flagellar,polotzek2013three,geyer2013cell,bennett2013emergent,bennett2013phase}.\ The relative importance of these mechanisms remains to be explored for the various asymmetric waveforms \cite{bennett2013phase} arising in our model.}

\begin{figure}[t]
	\centering
	\includegraphics[width=\linewidth]{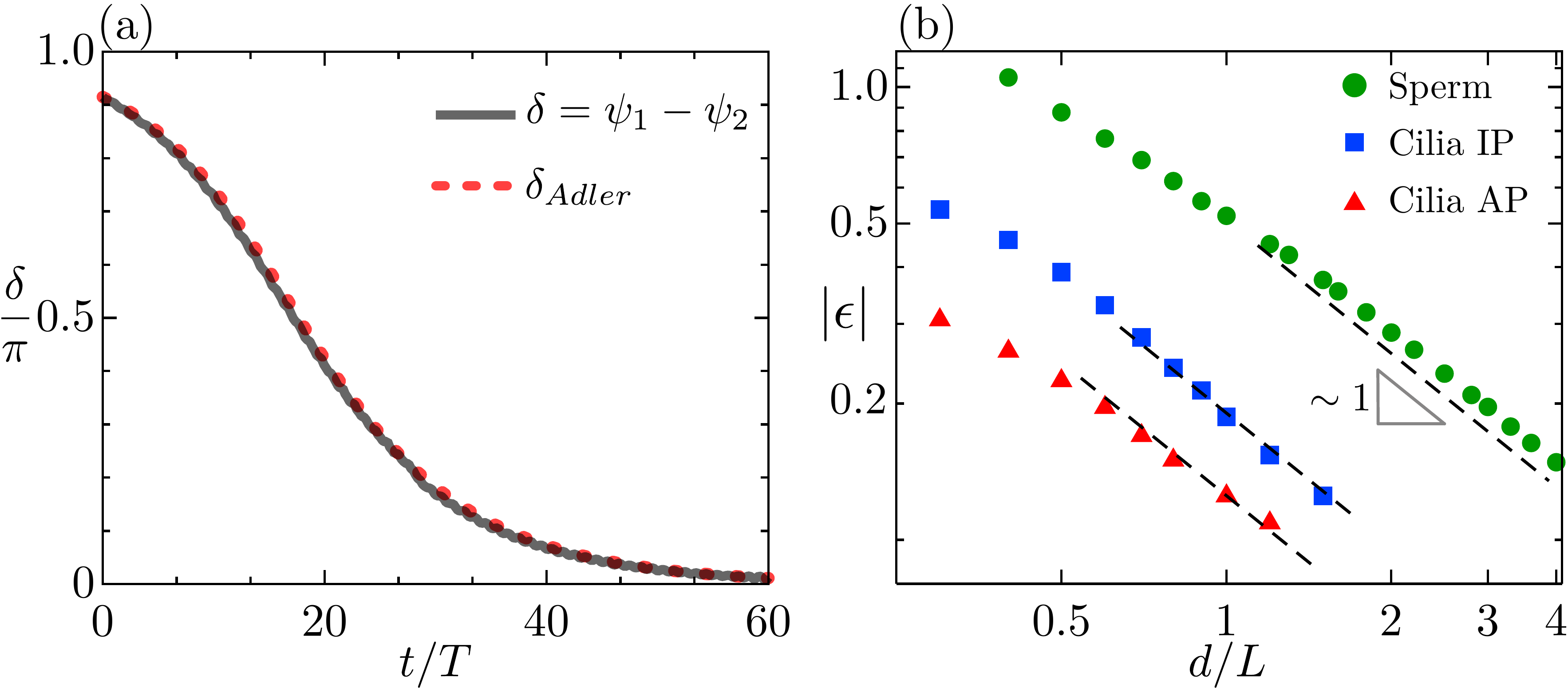}	\vspace*{-0.35cm}
	\caption{(a) Evolution of the phase difference $\delta(t)$ during synchronization of two nearby symmetric waveforms and comparison to the Adler equation (\ref{eq:adler}).  {Fitting parameter values: $\epsilon = -1.02$ and $\alpha = 0.15$, with a standard deviation of $\pm 0.01$}.\ (b) Coupling strength $|\epsilon|$ as a function of interflagellar distance $d$ for various beating patterns.}
	\label{fig:adler}
	\vspace*{-0.1cm}
\end{figure}

For a more quantitative analysis of synchronization, we introduce a definition of the phase $\psi$ of a waveform.\ To this end, we perform the Hilbert transform of the continuous periodic time series $\beta(t) = \phi(1/2,t)$, providing the analytic continuation $\smash{\zeta(t) = \beta(t) + \mathrm{i} \widehat{\beta}(t)}$ where $\smash{\widehat{\beta}(t) \equiv (1/\pi) \dashint_{-\infty}^{\infty} \beta(\tau)/(t-\tau)\,d \tau}$.\ The phase of the waveform is then calculcated as ${{\psi}(t) = \arctan[\widehat{\beta}(t)/\beta(t)]}$, and we use an appropriate geometric gauge to define a true phase that grows monotonically with time \cite{kralemann2008phase}.\ The phase difference $\delta(t) = \psi_1 - \psi_2$ for two nearby sperms going from AP to IP is shown in Fig.~\ref{fig:adler}(a) and decays to zero over the course of several periods.\ In spite of the complexity of the governing equations in presence of HI, the phase difference is well described by a simple low-dimensional Adler equation as in past experiments with \textit{Chlamydomonas}  \cite{goldstein2009noise} and in minimal rotor models \cite{niedermayer2008synchronization,uchida2011generic,maestro2018control,Note1}. Here, we seek a two-parameter equation of the form\vspace{-0.05cm}
\begin{equation}
\dot{\delta} = \epsilon \sin \delta + \alpha \sin 2 \delta, \vspace{-0.1cm} \label{eq:adler}
\end{equation}
where constants $\epsilon,\alpha$ are estimated numerically.\ A solution to this equation follows the numerical data very well in Fig.~\ref{fig:adler}(a). In all our computations, we find that $|\epsilon| \gg |\alpha|$ and thus define $|\epsilon|$ as the effective coupling strength in accordance with minimal models of rotors \cite{Note1}.\ When plotted as a function of  interflagellar distance $d$ in Fig.~\ref{fig:adler}(b),  {$|\epsilon|$ shows a far-field algebraic decay of $1/d$ over the limited range of accessible values, which is a signature of the dominant Stokeslet HI and can be rationalized from a simple rotor model \cite{Note1}. A slower decay is seen at short separations, where complex near-field interactions take place.}\ Stronger coupling arises for symmetric spermlike beats than for ciliary beats, primarily due to the longer lengths of sperm flagella.\ For cilia, we also find that $|\epsilon|_\mathrm{IP} > |\epsilon|_\mathrm{AP}$  in agreement with experiments \cite{brumley2014flagellar}, which can be attributed to the fact that filaments spend more time close to one another during IP beats and thus interact more strongly. 

\begin{figure}[t]
	\centering
	\includegraphics[width=\linewidth]{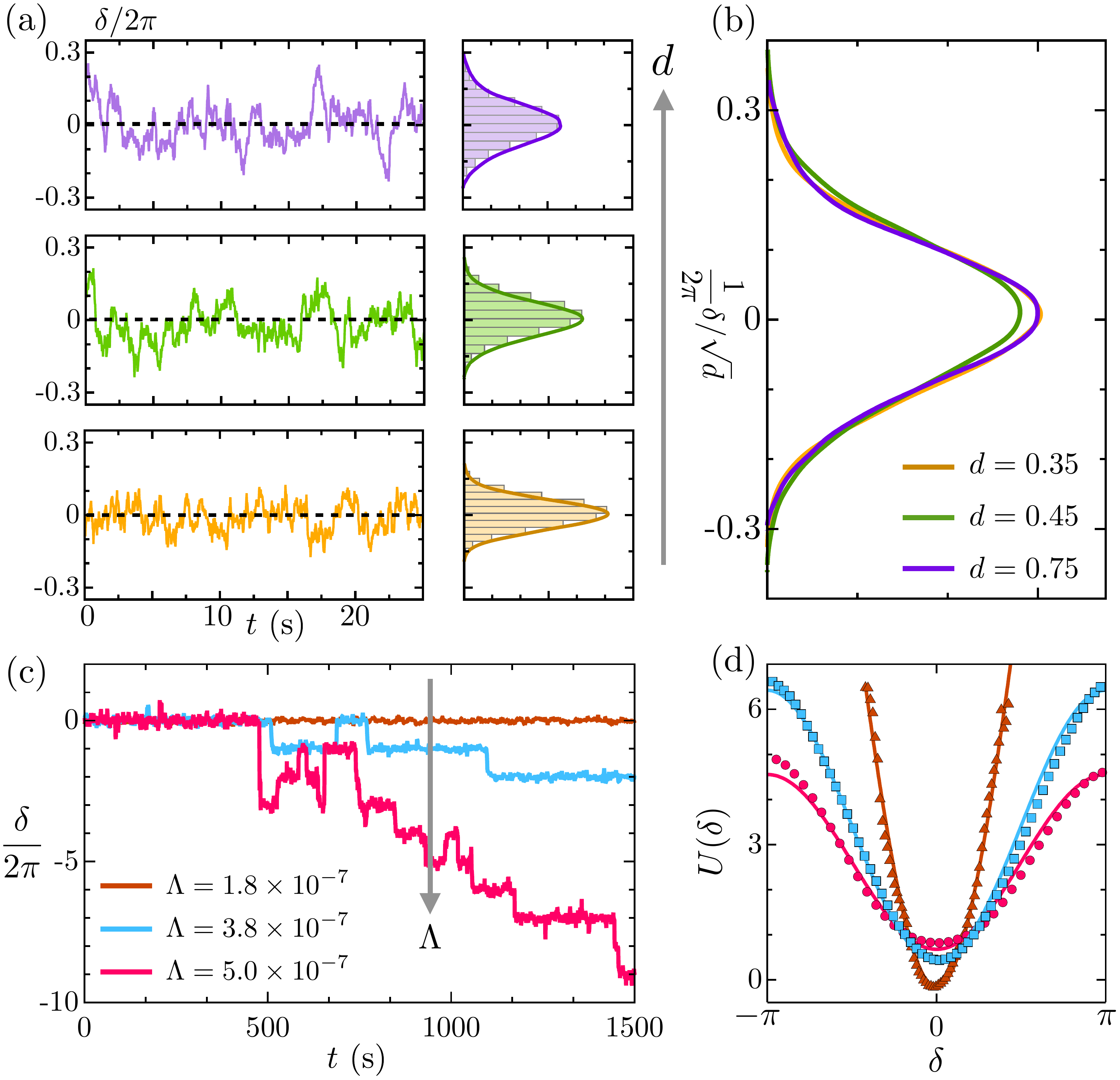}
	\vspace*{-1.5em}
	\caption{(a) Gaussian distributions of the fluctuations of the phase difference for varying separation distance $d$.\ (b) Collapse of the distributions in the rescaled variable $\smash{\delta/\sqrt{d}}$.\ (c) Long-time evolution of $\delta(t)$ for increasing biochemical noise $\Lambda$ at a fixed separation distance $d$, showing the emergence of slips. (d) Effective interaction potential $U(\delta)$ estimated from the statistics (symbols) and compared to the Fokker-Planck prediction (lines).\  {The diffusivity values for increasing noise levels are $D=0.11,0.33,0.58$.}}
	\label{fig:slips}
	\vspace*{-0.5em}
\end{figure}

Intrinsic to the kinetics of molecular motors is biochemical noise, which alters the precise notion of synchronization.\ To probe its effects, we study the long-time statistics of the phase difference in presence of noise for spermlike waveforms in Fig.~\ref{fig:slips}(a).\ Fluctuations follow a Gaussian distribution centered around the mean IP configuration of $\delta = 0$, with a variance scaling linearly with separation distance $d$.\ This is a consequence of the $1/d$ decay of the coupling strength $|\epsilon|$ and is further corroborated by the collapse of the distributions under the rescaling $\delta \to \delta/\sqrt{d}$ in Fig.\ \ref{fig:slips}(b) \cite{brumley2014flagellar}.\ We model the noisy phase dynamics by a stochastic Adler equation $\dot{\delta} = \epsilon \sin \delta + \chi(t)$ with $\langle \chi(t)\rangle=0$ and $\langle \chi(t) \chi(t')\rangle = 2 D \delta(t-t')$, where $D$ is the phase diffusivity with units of s$^{-1}$.\ Associated with the Adler equation is a Fokker-Planck description for the probability distribution $P(\delta)$ of the phase difference, with steady-state solution given by $P(\delta) = \exp(-\epsilon \cos \delta/D)/2 \pi I_0 (|\epsilon|/D)$, where $I_0$ is the modified Bessel function of order zero and  where we estimate $D$ numerically \cite{chakrabartispontaneous,stratonovich1967topics}.\ The interaction potential $U(\delta) = -\ln P(\delta)$, which is $2\pi$-periodic, is shown in Fig.\ \ref{fig:slips}(d) for increasing noise levels.\ When noise is weak, the filaments remain phase-locked and fluctuate around the IP configuration, which translates into a deep potential well at $\delta=0$.\ With increasing noise,  the potential well flattens as deviations from perfect AP synchrony become more frequent and  intense.\ Occasionally, accumulated noise allows the filaments to gather a complete phase of $2\pi$, causing them to `slip' towards $\delta\pm 2\pi$. These slips are visible in the phase trajectories of Fig.~\ref{fig:slips}(c) and can be interpreted as thermally assisted hops between neighboring wells in the flattened periodic potential. In absence of frequency mismatch, slips are equally probable in $\pm 2\pi$, and the stochastic Adler model predicts a frequency of $G = D |I_0 (|\epsilon|/D)|^{-2}/4 \pi^2$ \cite{stratonovich1967topics}.\  {Using the computed value of $D$, this prediction indeed provides a quantitative estimate of the mean frequency of slips in full simulations; see Supplemental Information \cite{Note1}.}


\textit{Concluding remarks.---} {We have used an idealized planar model of the flagellar axoneme that captures the essential physics of internal dynein activity and produces spontaneous oscillations similar to those seen in nature \cite{Note1} to elucidate elastohydrodynamic synchronization of nearby flagella and cilia.\ Our simulations underscore the essential roles of} hydrodynamic interactions and associated mechanochemical feedback in enabling synchronization.\ Our model predictions for various beating patterns and orientations all agree with experiments and give credence to  {a combination of sliding and curvature control mechanisms for the generation of spontaneous beats}.\ We were also able to reproduce experimentally observed phase slips induced by biochemical noise.\ Future studies with our model will probe the role of elastic basal couplings \cite{wan2016coordinated}, swimming cells that are free to adjust phase by sliding past one another \cite{yang2008cooperation} or  {by rotational motion of their body} \cite{geyer2013cell,bennett2013phase}, and emergent dynamics in large-scale ciliary arrays. \vspace*{-0.5cm}

\bibliography{bibfile}

\end{document}


\title{Supplemental Information}
\maketitle
\begin{center}
	\textbf{Hydrodynamic synchronization of spontaneously beating filaments} \\
	\vspace{0.1cm}Brato Chakrabarti and David Saintillan \\
	{\it Department of Mechanical and Aerospace Engineering,\\
		University of California San Diego, 9500 Gilman Drive, La Jolla, CA 92093, USA }
\end{center}

\section{Parameters for simulations}
We have used an idealized two-dimensional model of the flagellar axoneme that retains the essential physics of force generation by molecular motors as well as geometric feedback mechanisms leading to spontaneous oscillations. An extensive exploration of the model for isolated filaments can be found in \cite{chakrabartispontaneous}. Here we provide the range of key parameters used to obtain various waveforms that we used to study hydrodynamic interaction in the main text. Dimensional parameter values used to produce symmetric beats are provided in Table~\ref{tabdim}, while the range of the corresponding dimensionless groups is given in Table~\ref{tabnodim}.

\begin{table}[H]
	\centering
	\renewcommand{\arraystretch}{1.1}
	\begin{tabular}{|c|c|l|}
		\hline
		Parameter    & Numerical Value      & Description                            \\ \hline
		$a$          & $200\,$nm                                     & Effective diameter of the axoneme \cite{riedel2007molecular}                              \\
		$L$          & $50\,\mu$m                          & Length of human sperm flagellum \cite{gaffney2011mammalian}  \\
		$B$          & $0.9-1.7 \times 10^{-21}\,$N$\,$m$^2$ & Range of bending rigidity of sea-urchin sperm and bull sperm \cite{riedel2007molecular,bayly2014equations} \\ 
		$K$          & $2 \times 10^3\,$N$\,$m$^{-2}$ & Inter-doublet elastic resistance measured for \emph{Chlamydomonas} \cite{oriola2017nonlinear} \\
		$\xi_\perp$    & $10^{-3}-1\,$Pa$\,$s         &  Range of coefficient of normal drag in different viscous media \cite{oriola2017nonlinear,bayly2014equations} \\
		$f_0$        & $1-5\,$pN                & Stall force for motor dynamics \cite{oriola2017nonlinear} \\
		$f_c$        & $0.5-2.5\,$pN            & Characteristic unbinding force of the motors \cite{howard2001mechanics} \\
		$v_0$        & $5-7\,\mu$m$/$s           & Motor walking speed at zero load \cite{howard2001mechanics} \\
		$\tau_0$     & $50\,$ms                 & Correlation time of motor activity \cite{oriola2017nonlinear} \\
		$\rho$        & $10^3\,\mu$m$^{-1}$                  & Mean number density of motors \cite{oriola2017nonlinear} \\
		\hline
	\end{tabular}
	\caption{Table listing the numerical values of the dimensional parameters  as reported in various experiments. }
	\label{tabdim}
\end{table}

\begin{table}[H]
	\centering
	\renewcommand{\arraystretch}{1.1}
	\begin{tabular}{|c|c|}
		\hline
		Dimensionless number    & Numerical value  \\ \hline
		$\mathrm{Sp} \equiv L\left(8 \pi \nu/B\tau_0\right)^{1/4}$         & $5-20$         \\
		$\mu_a \equiv a f_0 \rho L^2/B$       & $2 - 25$ $\times 10^3$    \\
		$\mu \equiv K a^2 L^2/B$          & $50-100$  \\ 
		${f}^* \equiv f_0/f_c$ & 2    \\
		$\zeta \equiv a/v_0 \tau_0$ & $\mathcal{O}(1)$  \\ \hline
	\end{tabular}
	\caption{Range of the dimensionless groups of the problem estimated using the parameter values of Table~\ref{tabdim} .}
	\label{tabnodim}
\end{table}

Simulations of asymmetric waveforms were performed using slightly different values as listed here:
\begin{itemize}
\item Cilia-like beats: Cilia come in a variety of lengths in the range of $L \sim 4-20\,\mu$m \cite{dillon2000integrative} that are typically shorter than human sperm flagella.\ We choose a bending rigidity of $B \sim 5 \times 10^{-23}$\,N\,m$^2$, which is slightly larger than that reported in \cite{gueron1997cilia}. The characteristic time scale $\tau_0 \approx 40$\,ms is chosen to produce a dimensional beating frequency of $f \sim 10-20$\,Hz, which is typical of muco-ciliary transport \cite{delmotte2006ciliary}. With a choice of $L \approx 8\,\mu$m, we estimate the typical sperm number to be $\mathrm{Sp} \sim 2-3$. In order to have asymmetry in beating we have used biased motor kinetics. In particular, for waveforms in 
Fig.~1 of the main text we use:  $\mathrm{Sp}=2.5$, $\mu_a = 2\times 10^3$, $\mu = 90$, $(\pi_0^+,\epsilon_0^+) = (0.17,0.73)$ and $(\pi_0^-,\epsilon_0^-) = (0.25,0.75)$.

\item  \textit{Chlamydomonas reinhardtii} has a typical body size of $7-10\,\mu$m with two flagella of length $10-12\,\mu$m \cite{guasto2010oscillatory}.\ The bending rigidity of the axoneme is estimated to be $B \sim 5.8 \times 10^{-22}$\,N\,m$^2$. Our numerical experiments indicate that waveforms of the type shown in Fig.~1 in the main text are possible for $\mathrm{Sp}\sim 2-3$. Using a characteristic time scale of $\tau_0 \approx 40$\,ms, we observe a beating frequency $f\sim 12-15$\,Hz, which is lower than the typical frequency of $50-60$\,Hz observed in experiments. In particular, for \emph{Chlamydomonas} waveforms in Fig.~1 of the main text we use: 
$\mathrm{Sp}=2.2$, $\mu_a = 1.5\times 10^3$, $\mu = 80$, ${f}^* = 1.9$, $(\pi_0^+,\epsilon_0^+) = (0.24,0.76)$, $(\pi_0^-,\epsilon_0^-) = (0.39,0.61)$, $\kappa^0 = -2.0$ and $\kappa_c = 12$.
\end{itemize}

\section{Predicted waveforms and comparison with experiments}

\subsection{Sperm-like beating and nonlinear oscillations}
\begin{figure}[b]
	\centering
	\includegraphics[width=0.63\linewidth]{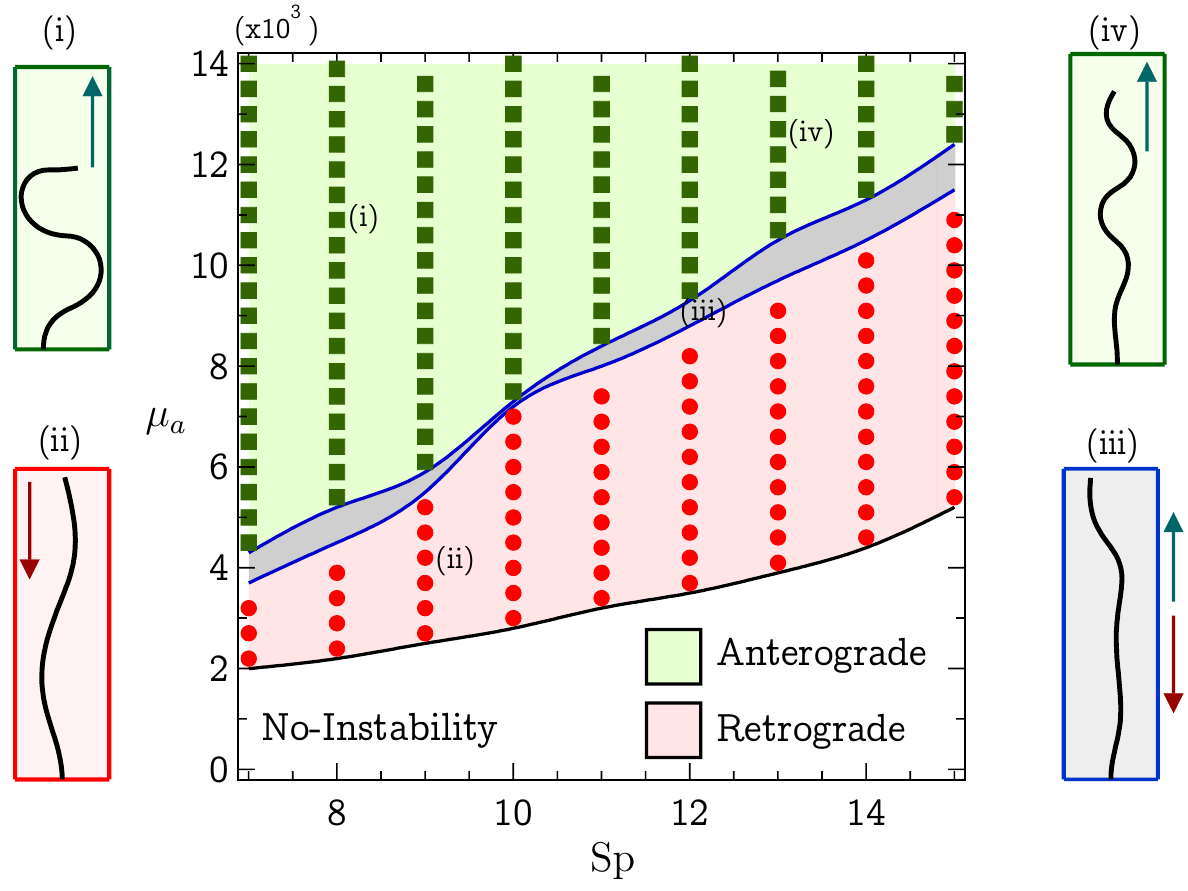}\vspace{-0.2cm}
	\caption{Phase chart highlighting the various dynamical transitions and beating patterns in the ($\mathrm{Sp},\mu_a$)  parameter space. Waveforms on both sides show characteristic filament conformations, where the arrows indicate the direction of wave propagation. Shown simulations are for $\mu = 100$, $\zeta = 0.3$ and  $\eta = 0.14$. Reproduced from \cite{chakrabartispontaneous} with permission.}
	\label{pchart}
\end{figure}
The present model was previously analyzed by Oriola \textit{et al.}~\cite{oriola2017nonlinear} in the geometrically linear regime of small curvature.\ A linear stability analysis reveals that for a given sperm number $\mathrm{Sp}$ there exists a critical activity level $\mu_a^{crit}$ above which the filament undergoes a Hopf bifurcation (see Fig.~\ref{pchart} for a phase diagram). Above the bifurcation, a linear mode becomes unstable and grows exponentially while oscillating with a characteristic frequency $f$.  Close to the bifurcation, the instability induces a traveling wave that propagates along the filament from tip to base (retrograde propagation).\ This direction of propagation is a common feature of linearized sliding control models with clamped boundary conditions \cite{camalet1999self,hilfinger2009nonlinear,bayly2015analysis,goldstein2016elastohydrodynamic} but is inconsistent with the well-known beating patterns of swimming spermatozoa, which exhibit anterograde wave propagation from base to tip.

Further away from the bifurcation, our numerical simulations of the fully nonlinear sliding-control model reveal a second transition to base-to-tip (anterograde) wave propagation \cite{chakrabartispontaneous}. This transition occurs at approximately twice the critical level of activity required to trigger the first onset of oscillations. The transition from retrograde to anterograde wave propagation with increasing motor activity occurs through a narrow transitional band of intermediate states (gray region in the phase chart of Fig.~\ref{pchart}), {whose width varies slightly with sperm number. Detailed analysis of the parameter space can be found in \cite{chakrabartispontaneous}. 
	
In order to compare the anterograde regime of nonlinear, finite amplitude wave-propagation to experimental waveforms, we report tangent angle dynamics and simulated snapshots of traveling waves in Fig.~\ref{fig:spermwave}. We find that by tuning the two primary control parameters, namely the Sperm number $\mathrm{Sp}$ and the activity number $\mu_a$, we are able to observe a variety of anterograde waveforms with different wavenumbers, amplitudes and frequencies.
\begin{figure}[t]
	\centering
	\includegraphics[width=1\linewidth]{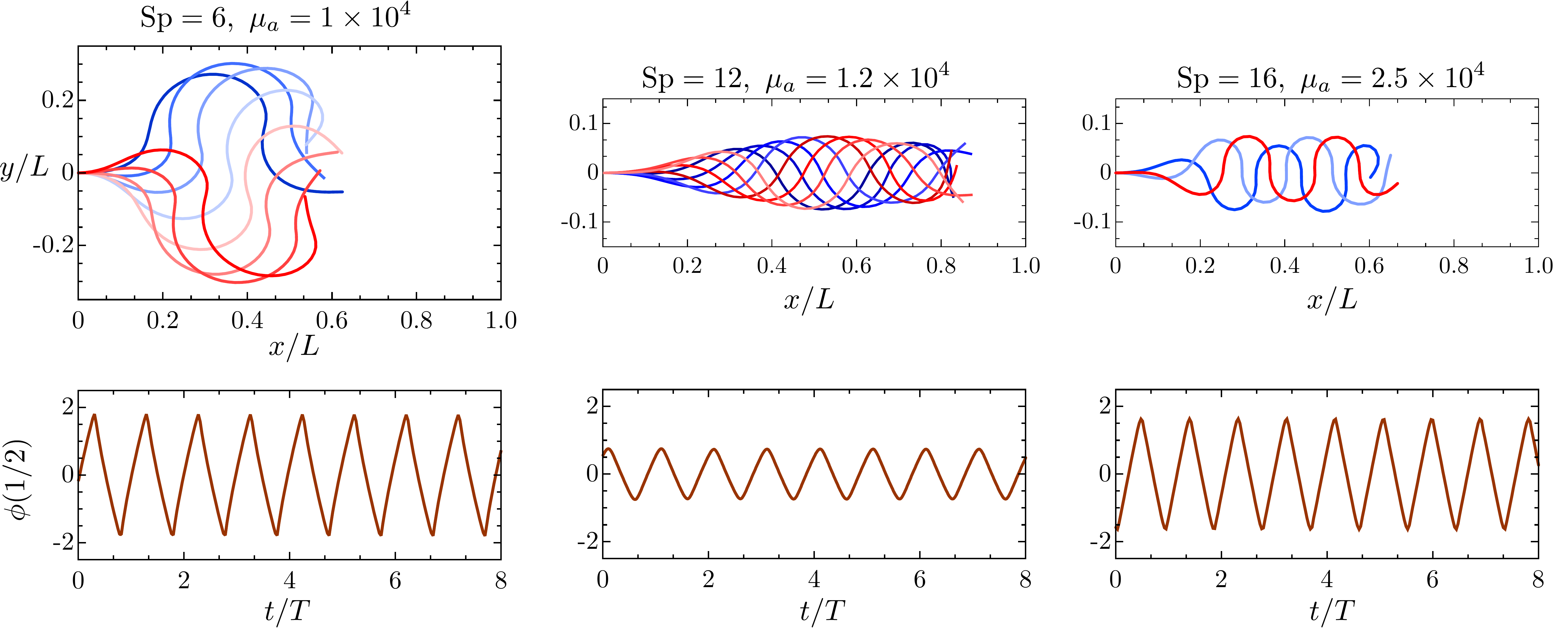}
	\caption{Sperm like waveforms for various combinations of $(\mathrm{Sp},\mu_a)$ and the corresponding evolution of the tangent angle at the filament midpoint. Simulations are for $\mu = 100$, $\zeta = 0.3$ and  $\eta = 0.14$.}
	\label{fig:spermwave}
\end{figure}
The above waveforms compare extremely well with the experimental snapshots of various marine spermatozoa reported in \cite{brokaw1965non} in different viscous media. In particular, we are able to capture a range of wavenumbers along the filament, consistent with experimental observations. The amplitude, frequency and wavelength of oscillations depend non-trivially on the control parameters and have been explored in detail in \cite{chakrabartispontaneous}. The evolution of the tangent angle dynamics also compares quantitatively with the measurements in \cite{riedel2007molecular} for bull sperm.  

More quantitative details on the waveform of Fig.~\ref{fig:spermwave}  can be obtained by analyzing the tangent angle dynamics. Following Sartori \textit{et al.}~\cite{sartori2016dynamic}, we present results on the power spectrum of the mean tangent angle, which is defined as $\smash{\langle \phi(t) \rangle_s = \int_0^L \phi(s,t)\,ds}$. Its power spectral density $\smash{\langle |\hat{\phi}|^2\rangle_s}$ is shown in Fig.~\ref{fig:power}, where we observe distinct peaks at the various harmonics of the fundamental frequency~$f$.\ For symmetric sperm-like beats, only odd harmonics appear in the spectrum, which is a direct consequence of the symmetry property $\phi(s,t+T/2) = -\phi(s,t)$ where $T=1/f$ is the period of oscillation. This property of the power spectrum has also been reported in \cite{sartori2019effect}. 
\begin{figure}[b]
	\centering
	\includegraphics[width=0.5\linewidth]{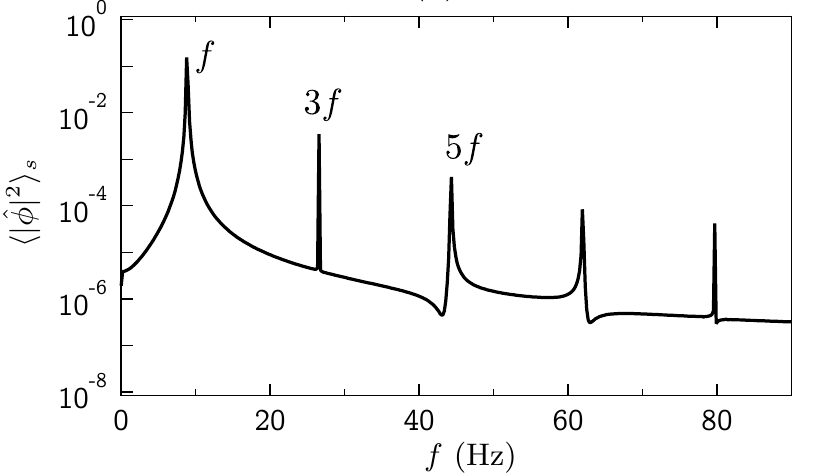}
	\caption{Power spectrum of the mean tangent angle for a symmetric sperm-like beat. Reproduced from \cite{chakrabartispontaneous} with permission.}
	\label{fig:power}
\end{figure}

\subsection{Waveforms of \textit{Chlamydomonas}}

We have previously pointed out that the frequency of oscillation predicted by our model of \textit{Chlamydomonas}-like beats is off compared to experimental values. Nonetheless, the qualitative features of the beats are consistent with observations. The power spectrum of the mean tangent angle shown in Fig.~\ref{fig:chpower} indeed agrees well with past experimental measurements \cite{sartori2016dynamic}.
\begin{figure}[t]
	\centering
	\includegraphics[width=0.5\linewidth]{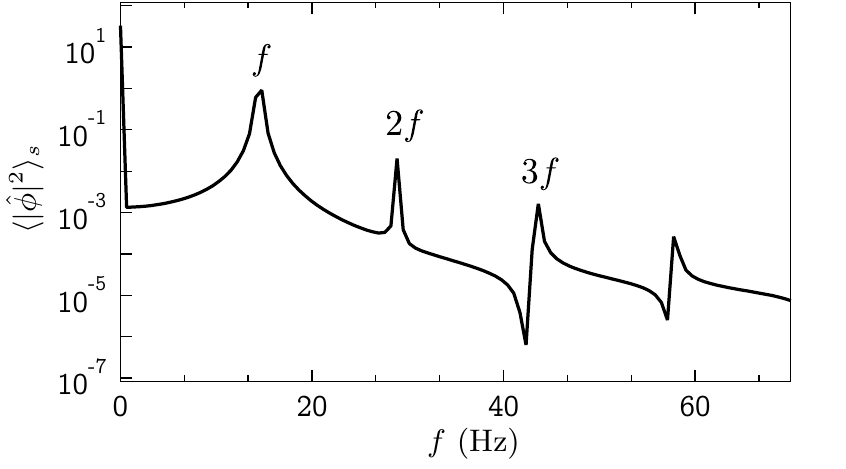}
	\caption{Power spectrum of the mean tangent angle for a \textit{Chlamydomonas}-like beat. Reproduced from \cite{chakrabartispontaneous} with permission.}
	\label{fig:chpower}
\end{figure}
In particular we find a peak at zero frequency corresponding to the non-zero mean shape arising from the static curvature of the filament. We also observe even harmonics of the fundamental frequency $f$ as the beating pattern is asymmetric. In order to further analyze the beating patterns, we followed \cite{geyer2013characterization,sartori2016dynamic} in seeking a Fourier decomposition of the tangent angle in the form
\begin{equation}
\phi(s,t) = \sum_{n=-\infty}^{\infty} \psi_n(s) \exp(\mathrm{i} n \omega t),
\end{equation}	
where $\omega$ is the angular frequency. In Fig.~\ref{fig:Fourier}, we show the static mode $\psi_0(s)$ and the first dynamic mode $\psi_1(s)$ and associated phase. In \cite{sartori2016dynamic}, this analysis was performed on a free axoneme, which is different from the clamped flagellum considered in our simulations. Nevertheless, our results compare favorably with the images analyzed in \cite{geyer2013characterization} (for example, figure 4.12 of the thesis) for a flagellum attached to the cell body of \emph{Chlamydomonas}. While there exist quantitative differences, our model is found to capture all the essential features of the beating pattern.
\begin{figure}[t]
	\centering
	\includegraphics[scale=0.28]{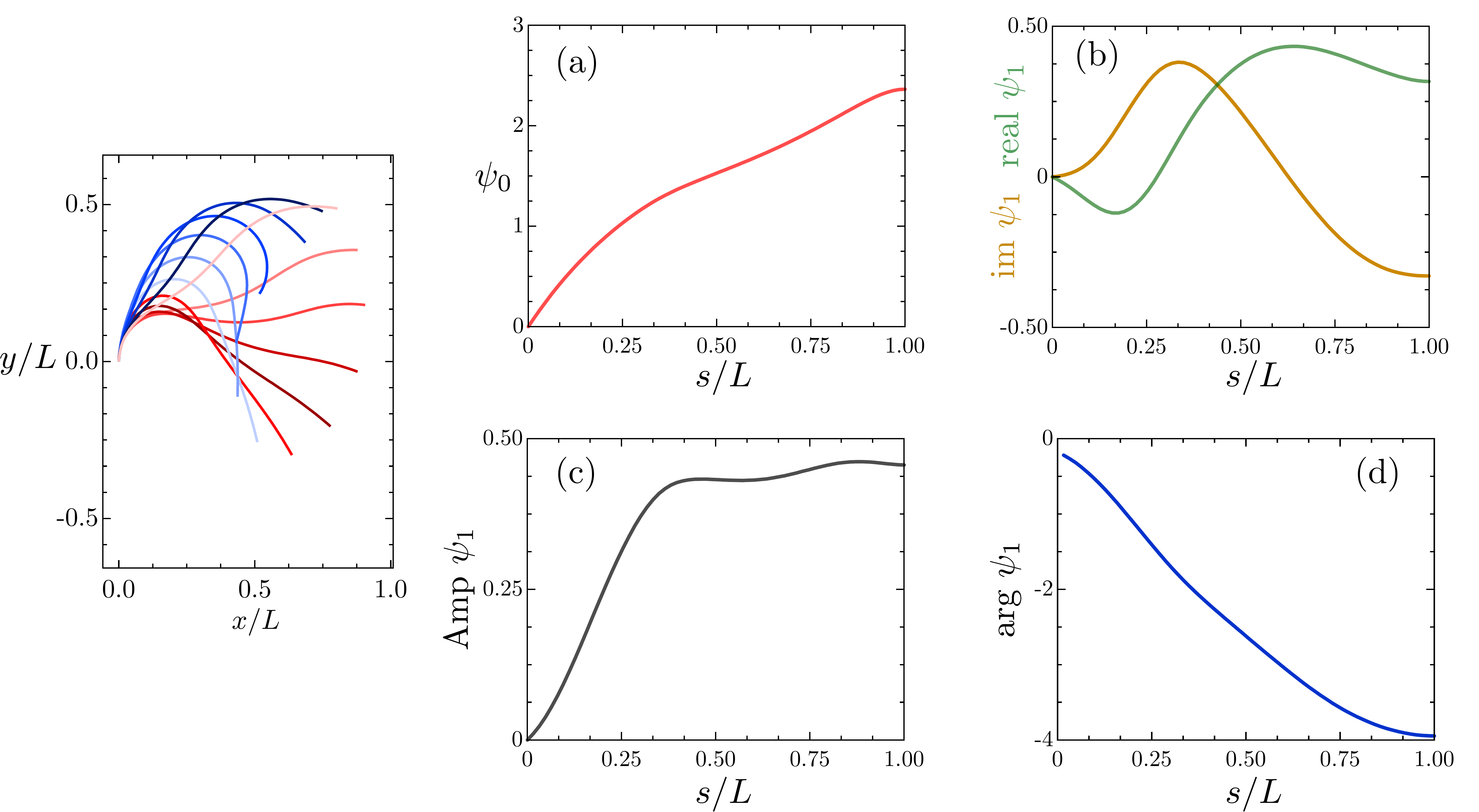}	
	\caption{Left: \textit{Chamydomonas}-like waveform. Right: Characterization of the waveform. (a) Zeroth mode $\psi_0$ of the beating pattern, showing the static curvature. (b) Real and imaginary parts of the first dynamic mode $\psi_1$. Since the flagellum is clamped, we have $\psi_1(0) = 0$, which is different from the free axoneme considered in~\cite{sartori2016dynamic}. (c) Amplitude of the first dynamic mode. (d) Argument of $\psi_1$, corresponding to the mode phase.}
	\label{fig:Fourier}
\end{figure}

\subsection{Ciliary beating patterns}

A wide variety of ciliary beating patterns are observed across biological systems. While most waveforms are three-dimensional \cite{blake1971spherical}, our present model and the associated formulation are restricted to planar beating patterns. We note, however, that cilia in a few biological organisms such as \emph{Opalina} \cite{sleigh1960form}, \emph{Ctenophore} \cite{gibbons1961relationship}, and molluscs \cite{gibbons1961relationship} are known to have almost planar beating patterns where our model can be relevant. Quite interestingly, defective respiratory cilia beat entirely in a plane \cite{chilvers2003ciliary} with a waveform similar to that shown in the main text. Fig.~\ref{fig:cilwave} shows two different waveforms that can be obtained with our model using various geometric feedback mechanisms.

We want to highlight that our focus in the main text has been to qualitatively capture the synchronization behavior observed in experiments with micropipette-held somatic cells of \emph{Volvox carteri} \cite{brumley2014flagellar}. The flagella in these cells beat almost entirely in a plane \cite{hoops1993flagellar}. Our computations show that in order to reproduce the synchronization behavior it is the asymmetry of the beating pattern that is crucial rather than the non-planarity. This behavior is consistent in our simulations with flagellar beating resembling \emph{Chlamydomonas}.
\begin{figure}[H]
	\centering
	\includegraphics[scale=0.35]{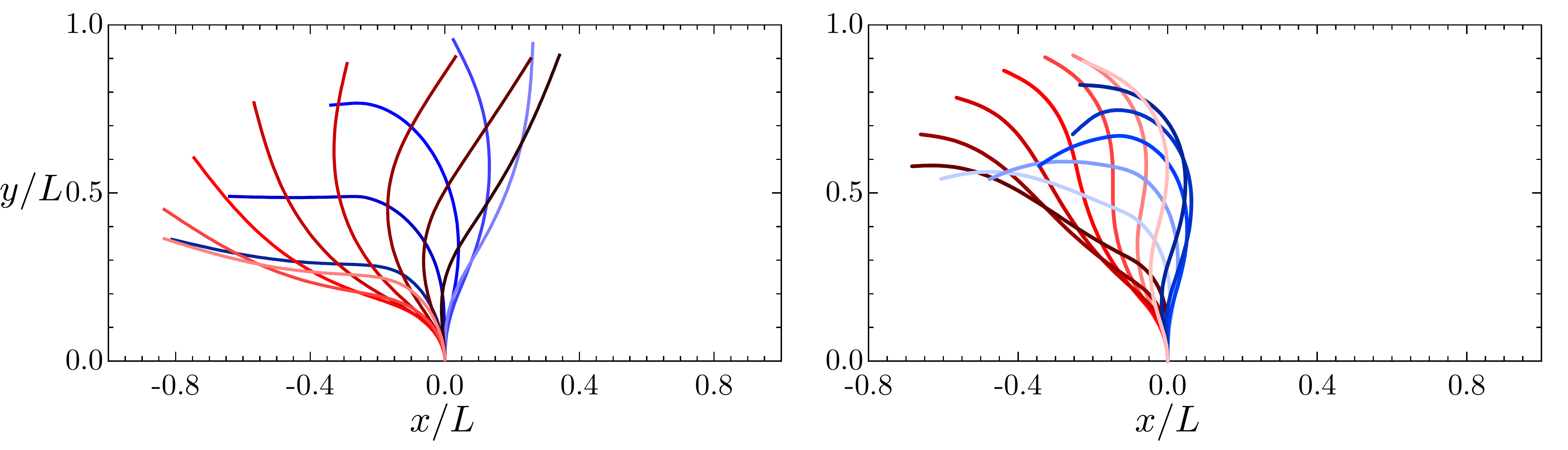}	
	\caption{Possible ciliary beating patterns under pure sliding control (left) and a combination of sliding and curvature control (right).}
	\label{fig:cilwave}
\end{figure}

\section{The Adler equation in a coarse-grained minimal model}

The main text highlights the complexity of the nonlinear governing equations of motion that incorporate hydrodynamic interactions. While it is challenging to systematically reduce the set of governing partial differential equations to a simple Adler equation, its role in capturing hydrodynamic synchronization can be appreciated from one of the first coarse-grained description of synchronization based on a  minimal model \cite{niedermayer2008synchronization}.
\begin{figure}[t]
	\centering
	\includegraphics[width=0.5\textwidth]{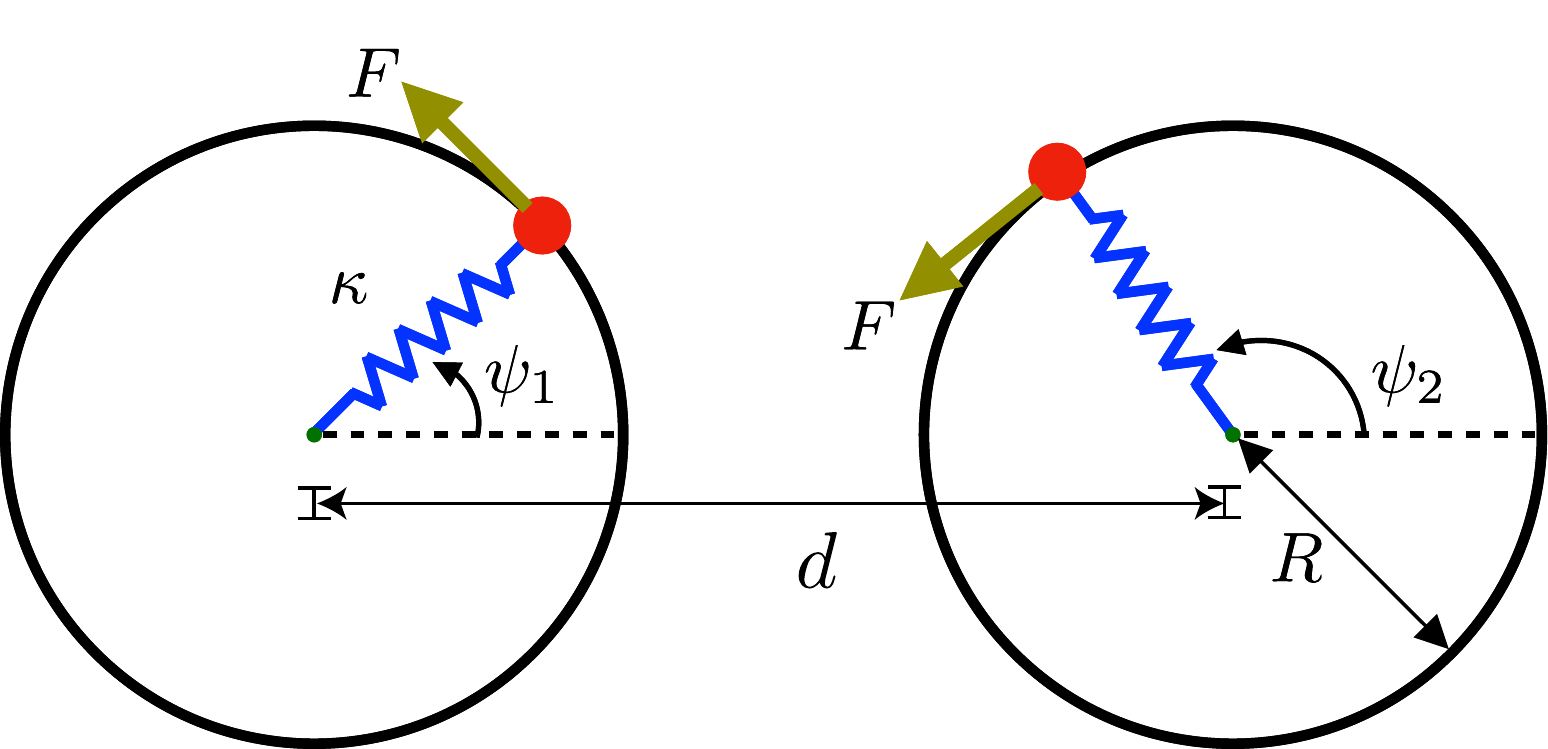}
	\caption{Schematic of the coarse-grained minimal model of Niedermayer \textit{et al.}~\cite{niedermayer2008synchronization}, in which the beating filaments are idealized as rotors moving along circular orbits.}
	\label{fig:force}
\end{figure}
Following Niedermayer \textit{et al.}~\cite{niedermayer2008synchronization}, one can idealize the flagella or cilia as beads driven on circular orbits with a compliance $\kappa$ capturing the role of elasticity. A constant force $F$ driving the beads represents the active forcing. The radius $R$ of the circular orbit is proportional to the length of the flagellum. In terms of the variables of the full model discussed in the main text, we can estimate $\kappa \sim B/L^3$ and $F \sim \rho f_0$ where $B$ is the bending rigidity of the flagellum, $\rho$ is the density of molecular motors and $f_0$ is the stall force. The friction coefficient $\zeta$ of the beads depends on the viscosity $\nu$ of the media. The radius $a$ of the micro-bead can be chosen to match the frequency of beating patterns. As the beads interact hydrodynamically, they are free to adjust their respective phase.  For sufficiently stiff springs and when the separation distance $d \gg L$, the phase difference between the beads can be shown to satisfy the simple ODE \cite{niedermayer2008synchronization}
\begin{equation}
\frac{\mathrm{d}}{\mathrm{d}t} (\psi_1-\psi_2) = \dot{\delta} = - \epsilon \sin \delta,
\end{equation}
where $\epsilon  \sim a \omega^2 \nu L^3/B d$ and $\omega$ is the angular frequency of rotation. We can relate the rotation rate to the driving force simply as $\omega = F/R \zeta$. This allows us to determine the scaling of the coupling constant in terms of the physical parameters of the problem:
\begin{equation}
\epsilon \sim \frac{a \rho^2 f_0^2 L}{\nu B d}.
\end{equation}
Indeed the above relation highlights the $1/d$ scaling recovered in the main text using our detailed model of the axoneme. The above expression was used in several experiments \cite{polin2009chlamydomonas,goldstein2009noise} to model the coupling between \emph{Chlamydomonas} flagella, where it is again well captured by the Adler equation for phase synchronization. The anti-phase synchronization for opposite power strokes corresponds to counter-rotation of the beads and is again consistent with this minimal picture.

Note that there has been a series of coarse-grained models that add further complexity to the simple model of \cite{niedermayer2008synchronization}. In particular, a coarse-grained driving force for a flagellum is not constant and should depend on phase $\psi$. A combination of compliance and phase-dependent force can show rich dynamics \cite{uchida2011generic,golestanian2011hydrodynamic,brumley2014flagellar,maestro2018control}. However, it is difficult to reduce the system down to a simple Adler equation in those cases. The fact that the solutions from the PDEs discussed in the main text are found to obey the Adler equation is non-intuitive, yet it ties in nicely with experimental observations and provides a simple framework for analyzing interactions. 

\section{Frequency of phase slips}

In the main text, we explained that as the noise floor in the motor kinetics is increased one starts to observe more frequent deviations from the phase-locked conformation of $\delta = 0$. This can lead to a phase slips resulting in $\delta \to \delta \pm 2\pi$. From the statistics of these slips, it is possible to determine the effective diffusivity numerically that allows one to estimate the frequency of slips. We compare in Fig.~\ref{fig:slips} the measured frequency from simulation to the analytical prediction obtained from a Fokker-Planck model with the associated diffusivity $D$ (see main text). Good agreement is found over the range of values accessible with our model. We note that the error in predicting the diffusivity increases with the noise level (larger values of $D$). 
\begin{figure}[t]
\centering
\includegraphics[scale=0.5]{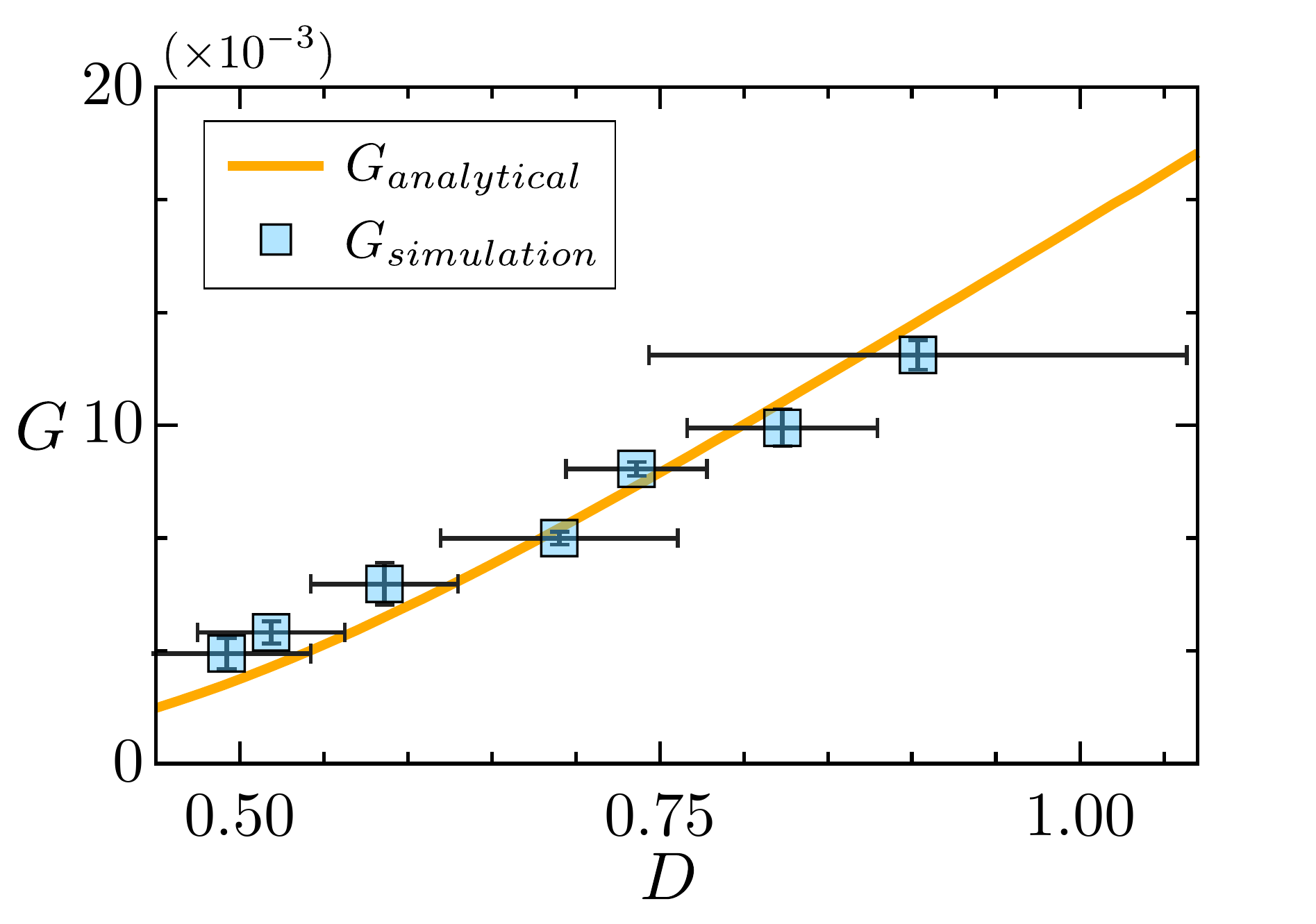}
\caption{Comparison of the measured frequency of phase slips $G$ in s$^{-1}$ to the analytical prediction from the Fokker-Planck model (see main text).}
\label{fig:slips}
\end{figure}

\bibliography{bibfile}